\documentclass[%
12pt,
  reprint,
 superscriptaddress,
 amsmath,amssymb,
 aps,UTF8,
]{revtex4-1}

\usepackage{graphicx}
\usepackage{dcolumn}
\usepackage{bm}
\usepackage{xcolor}
\usepackage{url}
\usepackage{float}
\usepackage{tabularx}
\usepackage{lipsum}
\usepackage{array}

\usepackage{booktabs}

\usepackage{subfigure}
\usepackage{multirow}

\begin{document}


\title{Mechanical structures inside proton with configurational entropy language}
\author{Wei Kou}
\email{kouwei@impcas.ac.cn}
\affiliation{Institute of Modern Physics, Chinese Academy of Sciences, Lanzhou 730000, China}
\affiliation{School of Nuclear Science and Technology, University of Chinese Academy of Sciences, Beijing 100049, China}
%

%
%
\author{Xurong Chen}
\email{xchen@impcas.ac.cn}
\affiliation{Institute of Modern Physics, Chinese Academy of Sciences, Lanzhou 730000, China}
\affiliation{School of Nuclear Science and Technology, University of Chinese Academy of Sciences, Beijing 100049, China}



\begin{abstract}
The structure of the proton remains a significant challenge within the field of Quantum Chromodynamics, with the origin of its spin and mass still lacking a satisfactory explanation. In this study, we utilize the gravitational form factor of the proton as the foundation for constructing the configurational entropy of the proton energy system. In our approach we choose the holographic QCD model for input thus obtaining a holographic version of the proton energy density. Employing this approach, we are able to determine key mechanical quantities such as the proton's mass radius and pressure distribution. Our analysis yields the root-mean-square mass radius of $\sqrt{\langle r_M^2\rangle}=0.720$ fm and scalar radius of $\sqrt{\langle r_S^2\rangle}=1.024$ fm for proton, which are found to be in excellent agreement with recent measurements from the Hall-C collaboration group at Jefferson Lab. Additionally, we examine the radial distribution of pressure and shear force within the proton. We provide a new mode for constraining holographic model parameters in the investigation of proton structures.
\end{abstract}

\maketitle


\section{Introduction}
\label{sec:intro}
For almost a century, physicists have been continuously fascinated by the discovery of the proton. In spite of the persisting efforts in recent decades, the enigma enveloping the proton has yet to be completely resolved. By now, we do have quite some information about the proton's radius, mass and spin. These characteristics stem from the intricate dynamics of the protons' basic constituents, quarks and gluons, which have been theoretically accounted for via Quantum Chromodynamic (QCD) theory \cite{Shifman:1978bx,Shifman:1978by,Shifman:1978zn}. Notably, the origin of proton mass still puzzles scientists, as the formation of a proton with a mass close to 1 GeV from a massless gluon and an almost massless quark appears unconventional. This emergence of mass in the proton seems to involve complex dynamical mechanisms that can be interpreted in ways other than through the Higgs mechanism \cite{Bernard:1995dp,DiVecchia:1980yfw,Witten:1980sp,Schwinger:1967tc}. Recent research on nucleon structure has focused on the study of mass decomposition, gravitational form factor of protons, and other related inquiries \cite{Goeke:2007fp,Teryaev:2016edw,Kharzeev:2021qkd,Mamo:2019mka,Mamo:2021krl,Mamo:2022eui,Hatta:2018sqd,Hatta:2018ina,Hatta:2019lxo,Pefkou:2021fni,Ji:2020bby,Ji:2021mtz,Ji:2021pys,Sun:2021gmi,Lorce:2021xku,Wang:2021dis,Wang:2021ujy,Yang:2018nqn,Shanahan:2018pib,Han:2022qet,He:2021bof,Kou:2021bez,Kou:2021qdc,Guo:2021ibg,Azizi:2019ytx,Azizi:2020jog,Ozdem:2019pkg,Ozdem:2022zig,Burkert:2023wzr}.

The precise measurement of the proton charge radius has provided insight into its electromagnetic nature \cite{Pohl:2010zza,A1:2010nsl,Antognini:2013txn,Mohr:2015ccw,Beyer:2017gug,Fleurbaey:2018fih,Xiong:2019umf,Tiesinga:2021myr}. Similar to the description of proton charge distribution through its charge radius, the mass radius of the proton can be explained by its mass distribution \cite{,Goeke:2007fp,Polyakov:2018zvc,Kou:2021bzs}. The heavy quarkonia photoproduction experiments near threshold have been proposed as a means to determine the mass radius of the proton \cite{Kharzeev:2021qkd,Wang:2021dis}. Recent GlueX and Hall-C Collaborations at Jefferson Lab have reported threshold data for photoproduction of charmonium J/$\psi$ \cite{GlueX:2019mkq,Duran:2022xag} that may provide insights into basic questions related to proton structure. In fact, the recent study described in \cite{Duran:2022xag} outlined the discovery of a shell-like outermost layer of the proton via its scalar radius. Separately, there is ongoing discussion between experimental and theoretical physicists regarding how best to extract the gravitational form factor of the proton - a key component in determining its mass radius - from experimental data. To this end, many researchers have employed holographic construction to determine the gravitational form factor of the proton \cite{Mamo:2019mka,Mamo:2021krl,Mamo:2022eui}, with a focus on obtaining the A-term and D-term contained therein. In addition, we note the information theory has also been explored as a tool to reveal the nucleon structure with various angles \cite{Kharzeev:2017qzs,Levin:2019fvb,Tu:2019ouv,Kharzeev:2021nzh,Kharzeev:2021yyf,Levin:2021sbe,Zhang:2021hra,Hentschinski:2021aux,Kou:2022dkw}. In this work, we propose the use of Shannon entropy in information theory to aid in obtaining input parameters for the proton gravitational form factor \cite{Mamo:2021krl,Mamo:2022eui}.

In the field of high-energy nuclear physics, information entropy has been explored in various areas such as hadron processes, lattice QCD, and AdS/QCD correspondence \cite{Ma:1999qp,Ma:2018wtw,Bernardini:2016hvx}. A noteworthy concept is the Shannon entropy, which serves as a novel measure for characterizing the information contained within the system's Hamiltonian and equations of motion. In functional space, this measure is known as configurational entropy (CE) \cite{Gleiser:2011di,Gleiser:2012tu,Gleiser:2015aav,Gleiser:2015rwa,Bernardini:2016hvx,Karapetyan:2016fai,daRocha:2022bnk,Colangelo:2018mrt}. As emphasized in Ref. \cite{Gleiser:2011di}, CE can offer insights into the localized nature of specific systems in high-energy physics. In this communication, we adopt CE, replacing the earlier use of Shannon entropy, to characterize the local spatial structure of the proton using a holographic model as input. Specifically, we investigate its impact on determining the proton's mass radius, pressure, and shear force distribution.
 
 
 The contents of this letter are organized as follows. In Sec. \ref{sec:formalism}, we briefly review relevant physics related to the proton gravitational form factor, as well as the CE methods employed in our calculations. This section also includes a description of the proton mass radius and associated pressure and shear distributions. In Sec. \ref{sec:resluts and discussion}, we present the complete calculations and corresponding discussions. Finally, in the last section, we summarize our findings and provide some concluding remarks and future outlook.

\section{Formalism}
\label{sec:formalism}

\subsection{Gravitational form factors in holographic construction}
\label{subsec:GFFs}
The three standard nucleon gravitational form factors can be defined by the full QCD energy-momentum tensor (EMT) \cite{Pagels:1966zza,Ji:1996ek,Polyakov:2018zvc}
\begin{equation}
	\label{eq:GFFs}
	\begin{aligned}
		&\left\langle p_2\left|T^{\mu \nu}(0)\right| p_1\right\rangle=  \bar{u}\left(p_2\right)\Big(A(K^2) \gamma^{(\mu} p^{\nu)}\\
		&+B(K^2) \frac{i p^{(\mu} \sigma^{\nu) \alpha} K_\alpha}{2 M_N}
		 +C(K^2) \frac{{K}^\mu K^\nu-\eta^{\mu \nu} K^2}{M_N}\Big) u\left(p_1\right),
	\end{aligned}
\end{equation}
with $a^{(\mu} b^{\nu)}=\frac{1}{2}(a^\mu b^\nu+a^\nu b^\mu)$, $K^\mu=p_2^\mu-p_1^\mu$, $p=\frac{1}{2}(p_1+p_2)$, where $T^{\mu\nu}=\sum_iT_i^{\mu\nu}$ with $i=g,q$ represents the full partons. All three $A(K^2$), $B(K^2)$ and $C(K^2)$ are scheme and scale-independent, similar to the electromagnetic Dirac and Pauli form factors, and therefore completely physical. The
normalization is $\bar{u}u=2M_N$. Eq. (\ref{eq:GFFs}) is conserved and tracefull. Throughout, $D(K^2)=4C(K^2)$ will be used interchangeably \cite{Mamo:2019mka}. In particular, throughout the letter we assume that $B(K^2)=0$ \cite{Pagels:1966zza,Ji:2021mtz}. These three gravitational form factors should be modeled as dipole or tripole form \cite{Pefkou:2021fni}. The corresponding parameters can be obtained from fitting the lattice data. In accordance with the findings presented in Refs. \cite{Mamo:2022eui,Pefkou:2021fni}, our study exclusively focuses on the gluon component of EMT and GFFs. This approach aligns with the concept proposed in Ref. \cite{Duran:2022xag}.


In order to determine the A-term $A(K^2)$, one needs to contract the EMT with the spin-2 transverse traceless polarization tensor $\epsilon_{\mu\nu}^{TT}$ mentioned in Ref. \cite{Mamo:2021krl} and uses a Witten diagram and the holographic dictionary in the soft wall construction to evaluate it \cite{Mamo:2019mka,Mamo:2021krl,Mamo:2022eui,Nastase:2007kj,Abidin:2009hr}. For the soft-wall model, the A form factor is written as \cite{Mamo:2021krl,Mamo:2022eui}
\begin{equation}
	\begin{aligned}
		\label{eq:A-term}
		&A(K^2)=  A(0)\Bigg[\left(1-2 a_K\right)\left(1+a_K^2\right) \\
		& +a_K\left(1+a_K\right)\left(1+2 a_K^2\right)\left(H\left(\frac{1+a_K}{2}\right)-H\left(\frac{a_K}{2}\right)\right)\Bigg],
	\end{aligned}
\end{equation}
with $a_K=\frac{K^2}{8\lambda_N}$. Here $H(x)=\psi(1+x)+\gamma_E$ is the harmonic number or digamma function plus Euler number. The scale $\lambda_N$ follows from the dilaton profile $\phi(z)=\lambda_N^2z^2$. In the large $N_c$ limit, $A(0)=1+\mathcal{O}(1/N_c)$ as the nucleon mass is totally glue dominated at low resolution. At finite $N_c$, only a fraction is glue dominated at the same resolution, so $A(0)$ is a free parameter that can be fixed by comparison to lattice results or experimental data \cite{Mamo:2022eui,Pefkou:2021fni}. In holographic QCD, the parameter $\lambda_N$ is conventionally determined via the $\rho$ meson Regge trajectory as $\lambda_N=m_\rho/2$ based on the soft-wall model. Recent studies \cite{Mamo:2019mka,Mamo:2021jhj,Mamo:2021krl,Mamo:2022eui} propose that the range of $\lambda_N$ should lie between approximately 330 MeV to 402 MeV, providing an accurate description of proton charge radius world data.

In this letter, we present a novel CE method for determining the parameter $\lambda_N$ in the context of physics. Our method is founded on the fundamental principle that the stability of a proton arises from the inherent selection process within the localized system. \cite{Karapetyan:2016fai,Karapetyan:2018oye,Karapetyan:2018yhm,Karapetyan:2021crv,Karapetyan:2021vyh}. Please note that the degrees of freedom of the system utilized in constructing the CE must be determined based on the subject of investigation. Hence, the ultimate outcome relies on the chosen model, exemplified by the holographic model examined in this work. To simplify our analysis, we adopt the assumption that the D form factor satisfies $D(K^2)=-4A(K^2)$ within the soft-wall model \cite{Mamo:2019mka}. We emphasize that this approximation differs from the treatment of the D form factor in Ref. \cite{Mamo:2022eui}.

The energy density $T_{00}(r)$ corresponding to the EMT is given from the determination of the individual gravitational form factors, which represents the eigenvalue of the Hamiltonian of the system \cite{Polyakov:2018zvc},
\begin{equation}
	\begin{aligned}
	\label{eq:energy density}
	T^{00}(r)&=\rho(r)=M_N\int \frac{d^3K}{(2\pi)^3}\exp(-iKr)\\
	&\times\left[A(K^2)-\frac{K^2}{4M_N}(A(K^2)-2J(K^2)+D(K^2))\right],
	\end{aligned}
\end{equation}
where $J(K^2)=\frac{1}{2}(A(K^2)+B(K^2))$ describe the spin-$1/2$ particle spin information \cite{Polyakov:2018zvc}. We emphasize that the expression (\ref{eq:energy density}) is analog to the electric charge distribution which can be mapped out by means of electron scattering experiments.
In an analog way, (hypothetical) scattering off gravitons would allow one to access
information on the spatial distribution of the energy inside a hadron \cite{Polyakov:2018zvc}. The purpose of energy density will be discussed in detail later.

\subsection{Radii, pressure and shear distributions}
\label{subsec:mechenics}

The determination of the gravitational A and D form factors allows for derivation of various radii of proton. Expanding on the research presented in Ref. \cite{Goeke:2007fp,Polyakov:2018zvc,Ji:2021mtz}, a comprehensive investigation of the gravitational form factor can be conducted. Specifically, the scalar or dilatation form factor can be defined via the trace part of (\ref{eq:GFFs}) in one such methodological approach,
\begin{equation}
	\label{eq:trace part}
	\langle p_2|T_\mu^\mu|p_1\rangle=\bar{u}(p_2)u(p_1) G_S(K^2),
\end{equation}
where 
\begin{equation}
	\label{eq:Gs}
	G_S(K^2)=A(K^2)-\frac{k^2}{4M_N^2}B(K^2)+\frac{3k^2}{4M_N^2}D(K^2).
\end{equation}
On the other hand, the (00) component defines the mass form factor in
the Breit frame,
\begin{equation}
	\label{eq:GM}
	G_M(K^2)=A(K^2)-\frac{k^2}{4M_N^2}B(K^2)+\frac{k^2}{4M_N^2}D(K^2).
\end{equation}
Following the definition of the electromagnetic radius of proton given by analogy with the electromagnetic Dirac and Pauli form factors, one can naturally obtain the radii definition corresponding to the above two form factors (\ref{eq:Gs},\ref{eq:GM}),
\begin{equation}
	\label{eq:radii}
	\begin{aligned}
		\langle r_S^2\rangle&=-6\frac{dG_S(K^2)}{dK^2}\bigg|_{K^2\to0},\\
		 \langle r_M^2\rangle&=-6\frac{dG_M(K^2)}{dK^2}\bigg|_{K^2\to0}.
	\end{aligned}
\end{equation}

The internal structure of the proton has been a subject of investigation in high-energy physics, and QCD has provided a framework for recognizing this structure. Although the constituents inside the proton are not yet fully understood, modern particle physics suggests that they consist of quarks and gluons, which interact through complex dynamics. However, explaining the distribution of pressure and shear force within the proton from first principles remains a challenge. The static EMT provides insight into this phenomenon, as the 
$ij$ component of the tensor defines the stress tensor. Further analysis has shown that the total stress tensor can be decomposed into two parts: a traceless portion associated with the shear force distribution $s(r)$, and a trace associated with the pressure distribution $p(r)$ \cite{Polyakov:2002yz,Polyakov:2018zvc}, 
\begin{equation}
	\label{eq:Tij}
	T^{ij}(\vec{r})=\frac13\delta^{ij}p(r)+\bigg(\hat{r}^i\hat{r}^j-\frac13\delta^{ij}\bigg)s(r),
\end{equation}
where $p(r)$ and $s(r)$ can be defined from the term of Fourier transform of $D(K^2)$,
\begin{equation}
	\label{eq:pressure}
	\begin{aligned}
		s(r)& =-\frac{1}{4M_N}r\frac{d}{dr}\left(\frac{1}{r}\frac{d}{dr}\tilde{D}(r)\right),  \\
		p(r)& =\frac1{6M_N}\frac1{r^2}\frac d{dr}\left(r^2\frac d{dr}\tilde D(r)\right),  \\
		\tilde{D}(r)& =\int\frac{d^3K}{(2\pi)^3}e^{-iKr}D(K^2). 
	\end{aligned}
\end{equation}
Note that the conservation of the EMT leads to $p(r)$ that satisfies the Laue condition \cite{Laue:1911lrk}, and this is necessary for the stability of the proton structure, but not sufficient.

\subsection{Configurational entropy and proton energy density}
\label{subsec:CE}
It is necessary to first briefly introduce the concept of CE. We use the description in Ref. \cite{Gleiser:2011di} as the beginning of the introduction to CE. The CE, based on
the Shannon’s information theory, comprises a procedure that logarithmically measures the underlying information of quadratically
Lebesgue-integrable functions, denoted hereon by $\rho(r)$ to further
represent the energy density that underlies the system to be analyzed, defined on $\mathbb{R}^d$. The energy density corresponding to the momentum space can be obtained from the Fourier transform
\begin{equation}
	\label{eq:rhok}
	\rho(K)=\int_{\mathbb{R}^d}d^d r\rho(r)e^{-iKr},
\end{equation}
which is the main ingredient to construct the modal fraction \cite{Gleiser:2012tu}
\begin{equation}
	\label{eq:modal}
	{f}(K)=\frac{|\rho(K)|^2}{\int_{\mathbb{R}^d}d^d K|\rho(K)|^2},
\end{equation}
which represents the weight of every single mode tagged by $k$. The CE associated with the system energy density is written as \cite{Gleiser:2011di,Gleiser:2012tu}
\begin{equation}
	\label{eq:CE}
	S_{CE}[\rho]=-\int_{\mathbb{R}^d}d^d K\tilde{f}(K)\log\tilde{f}(K),
\end{equation}
where $\tilde{f}(K)=f(K)/f(K)_{max}$ and $f(K)_{max}$ is the maximum fraction, in most cases of interest given by the zero mode, $K=0$ \cite{Gleiser:2011di}. The integrand $\tilde{f}(K)\log\tilde{f}(K)$ is called the CE density. Eq. (\ref{eq:CE}) is a continuous generalization of the original definition of Shannon's entropy $S=-\sum_k p_k\text{log}p_k $ \cite{Shannon:1948zz}. 

Critical points of the CE imply that the system has informational entropy that is critical with respect to the maximal entropy $\tilde{f}(k)_{max}$, corresponding
to more dominant states \cite{Bernardini:2016hvx,Casadio:2016aum,Karapetyan:2016fai}. It has been comprehensively reviewed and applied in various models, from high spin mesons and glueballs stability \cite{Bernardini:2016hvx,Bernardini:2016qit}. In addition to holographic nuclear physics and phenomenology there are also Bose-Einstein condensation and black hole related studies \cite{Casadio:2016aum,Braga:2016wzx}. The identical thinking is applied to the present work, as detailed in the next section.

\section{Results and discussions}
\label{sec:resluts and discussion}
In this section, we present the results of our numerical calculations and outline the overarching objective of this study. Our primary aim is to apply Eqs. (\ref{eq:rhok}-\ref{eq:CE}) for the purpose of computing the CE that corresponds to the energy density of the system (\ref{eq:energy density}). Specifically, we seek to determine the critical point in order to ascertain the parameter of the holographic version of the gravitational form factor (\ref{eq:A-term}). It should be noted that we assume $D(K^2)=-4A(K^2)$ and choose an integration dimension of $
d=3$ in the Breit frame \cite{Polyakov:2018zvc}. Subsequently, we employ the CE method to obtain the parameter $\lambda_N$, which can be utilized to calculate the proton mass/scalar radii (\ref{eq:radii}) and even the radial distribution of pressure and shear force (\ref{eq:pressure}).

\begin{figure}[htbp]
	\centering  
	\includegraphics[width=0.46\textwidth]{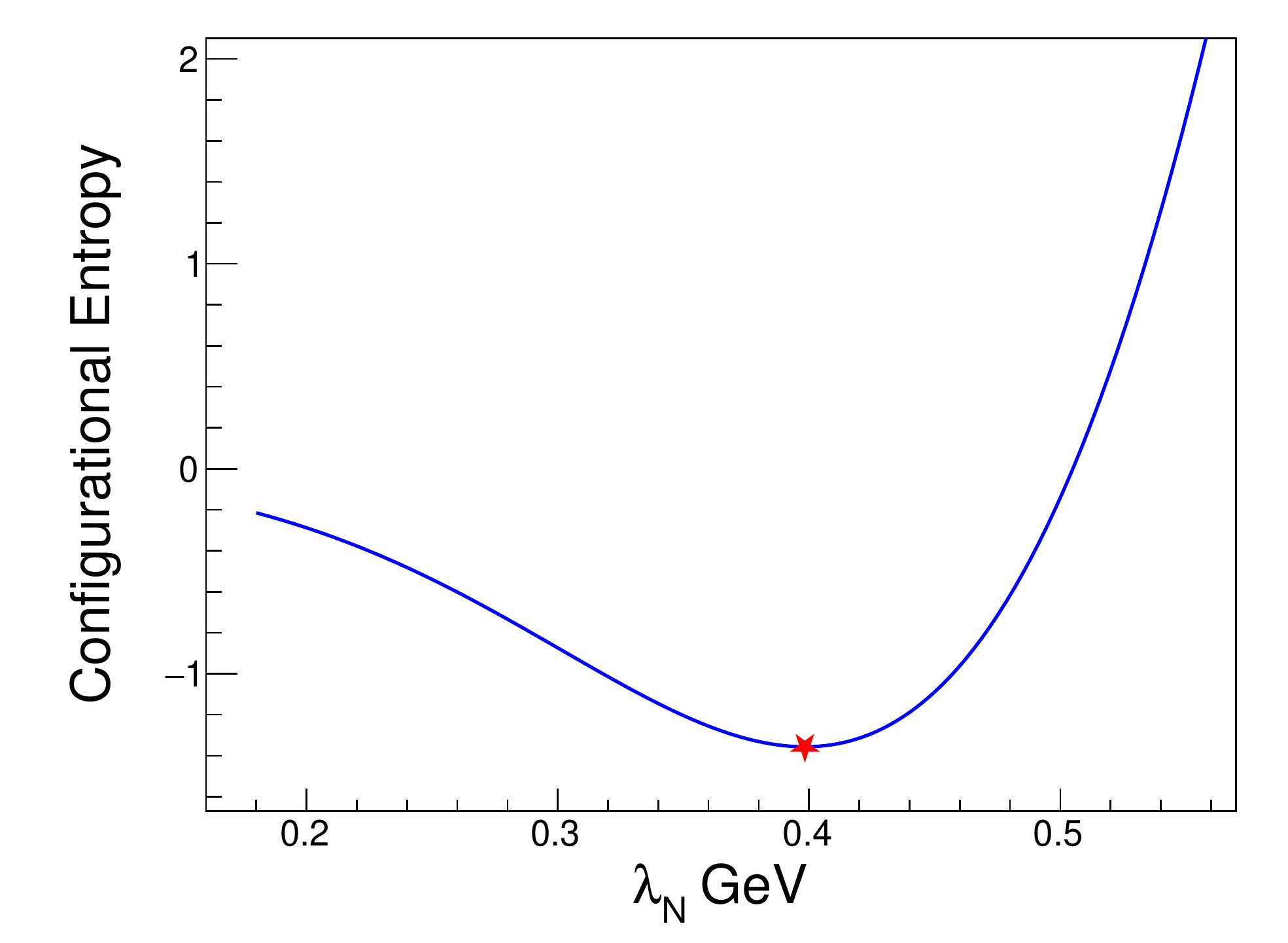}
	\caption{Configurational entropy (blue solid curve) as a function of the scale $\lambda_N$ mentioned above Eqs. (\ref{eq:A-term},\ref{eq:energy density}). The critical point is read as $\lambda_N=0.398$ GeV, $S_{CE}=-1.356$ nat (red star). }
	\label{fig:CE}
\end{figure}

Figure. \ref{fig:CE} displays the CE of proton system with QCD EMT and energy density. The critical point is at the minimum of our selection of the parameter space range, which is read as $\lambda_N=0.398$ GeV, $S_{CE}=-1.356$ nat. In Ref. \cite{Mamo:2022eui} the author provided the parameter $\lambda_N=0.388$ GeV, the results we obtained with the CE method differed from it by about 2.5$\%$. Note that the establishment of the CE method requires the use of a holographic model as input in the form of proton energy density. In addition, the parameter we obtained does not indicate an advantage over the parameter determined with half of the $\rho$ meson mass, we just provide an alternative way to roughly determine the model parameter. This attempt has been successful in condensed matter physics as well as gravitational cosmology, and we use it to consider the internal structure of the proton.

\begin{figure*}[htbp]
	\centering
	\subfigure{
		\includegraphics[width=0.46\textwidth]{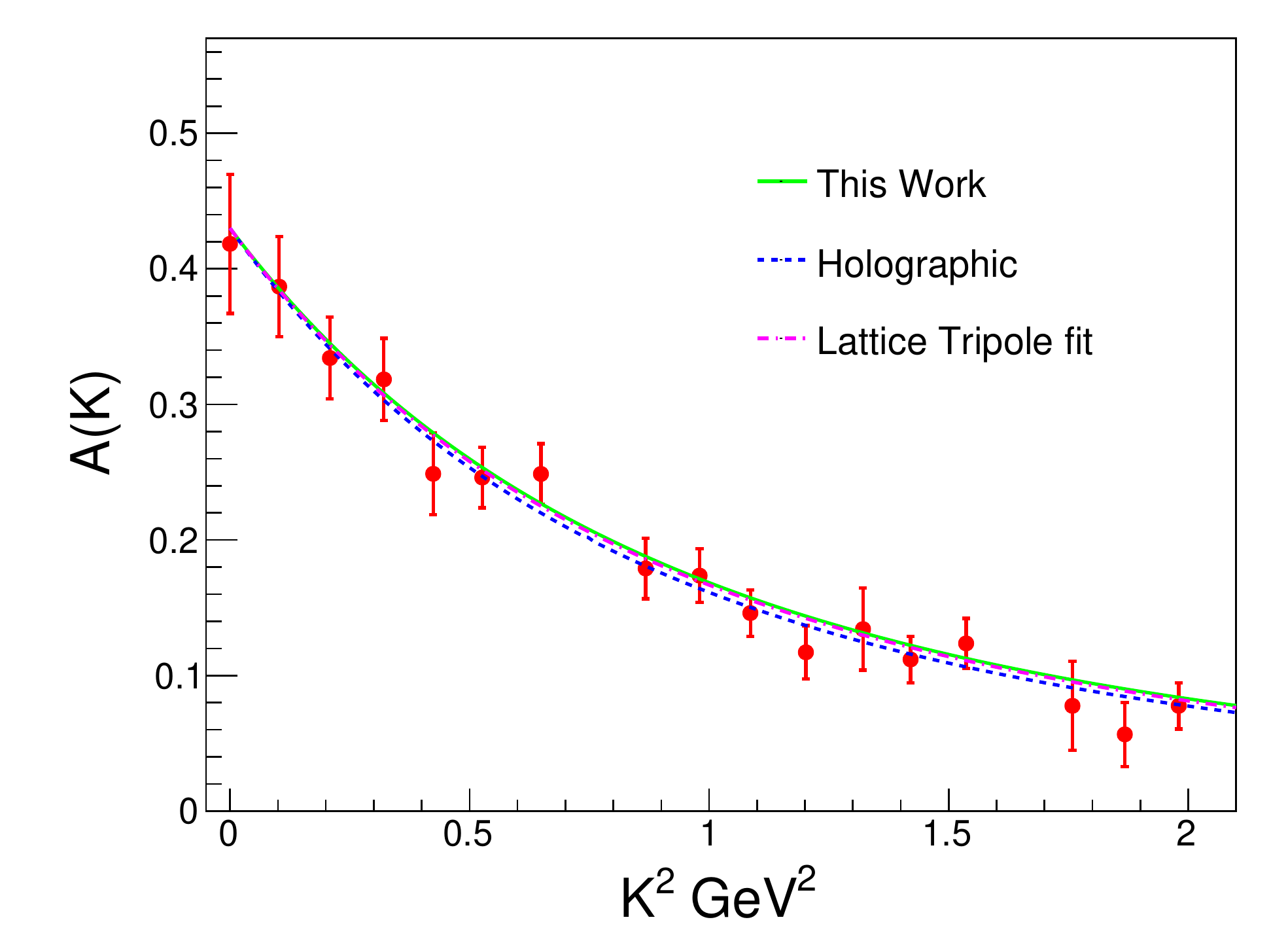}}
	\subfigure{
		\includegraphics[width=0.46\textwidth]{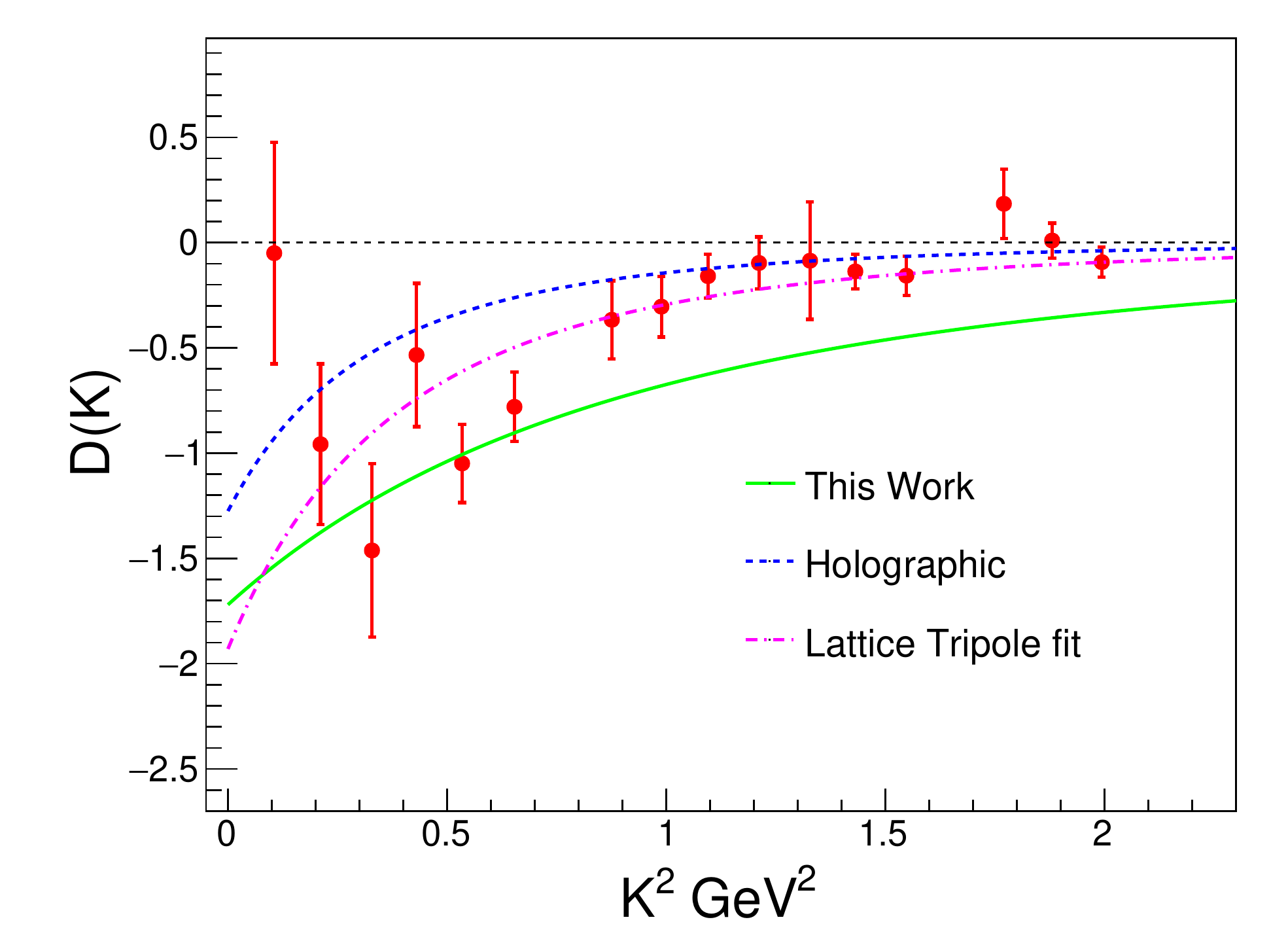}}
	\caption{(Left) The A-form factor from the recent lattice QCD result \cite{Pefkou:2021fni} (red point) and tripole form fit (magenta dashed-dotted curve), and the holographic fit using (\ref{eq:A-term}) with
		$\lambda_N=$0.388 GeV, and $A(0)=0.430$ \cite{Mamo:2022eui} (blue dashed curve). Our result is the green solid curve by using CE method, which gives the parameter $\lambda_N=0.398$ GeV. (Right) The D-form factor from the recent lattice QCD result \cite{Pefkou:2021fni} (red point) and tripole form fit (magenta dashed-dotted curve), our result (green solid curve) corresponds to $D(K^2)=-4A(K^2)$ and is distinct from the result from Ref. \cite{Mamo:2022eui} (blue dashed curve).}
	\label{fig:GFFs}
\end{figure*}

We present our findings on the A and D-form factors as depicted in FIG. \ref{fig:GFFs}. Our results are extracted via the CE method, although the normalization factor $A(0)$ is yet to be determined accurately due to its technical features. Thus, we have employed the fit of lattice data from \cite{Pefkou:2021fni} for compensation. Notably, we observe a well-matched A form factor calculated through CE (green solid curve) with the lattice data, which is also compared with the results from Refs. \cite{Mamo:2022eui,Pefkou:2021fni}. Meanwhile, the right panel of FIG. \ref{fig:GFFs} shows the contrastive outcomes for the D-form factor. This can be attributed to our utilization of the approximation $D(K^2)=-4A(K^2)$ when determining the parameters via the CE approach. Specifically, we apply the same set of parameters $A(0)=0.430,\ \lambda_N=0.398$ GeV to both the A and D-form factors.

The proton mass and scalar radii is calculated by Eqs. (\ref{eq:Gs}-\ref{eq:radii}). We compare the mass radius and scalar radius obtained by the CE method with the results reviewed in Refs. \cite{Mamo:2022eui,Pefkou:2021fni}, see Table. \ref{tab:radius}. The recent experimental extraction results can be found in Ref. \cite{Duran:2022xag}. One can find our calculated radii in the same order of magnitude as the results obtained by other methods. It suggests that it is possible to determine the proton gravitational form factor using the CE method.

\begin{table}[htbp]
	\caption{The proton mass radius $\sqrt{\langle r_M^2\rangle}$ and scalar radius $\sqrt{\langle r_S^2\rangle}$ are shown
		according to Eq (\ref{eq:radii}) with CE, holographic and lattice approaches \cite{Mamo:2022eui,Pefkou:2021fni} without displaying the uncertainty.}
	\label{tab:radius}
	\setlength{\tabcolsep}{2mm}{
		\begin{ruledtabular}
			\begin{tabular}{ccc}
				 Method& $\sqrt{\langle r_M^2\rangle}$ (fm)& $\sqrt{\langle r_S^2\rangle}$ (fm)\\ \hline
				 CE (This work)& 0.720 & 1.024\\
				 Holographic QCD & 0.682 &0.926\\ 
				 Lattice &0.746&1.073\\
			\end{tabular}
	\end{ruledtabular}}
\end{table}

\begin{figure*}[htbp]
	\centering
	\subfigure{
		\includegraphics[width=0.46\textwidth]{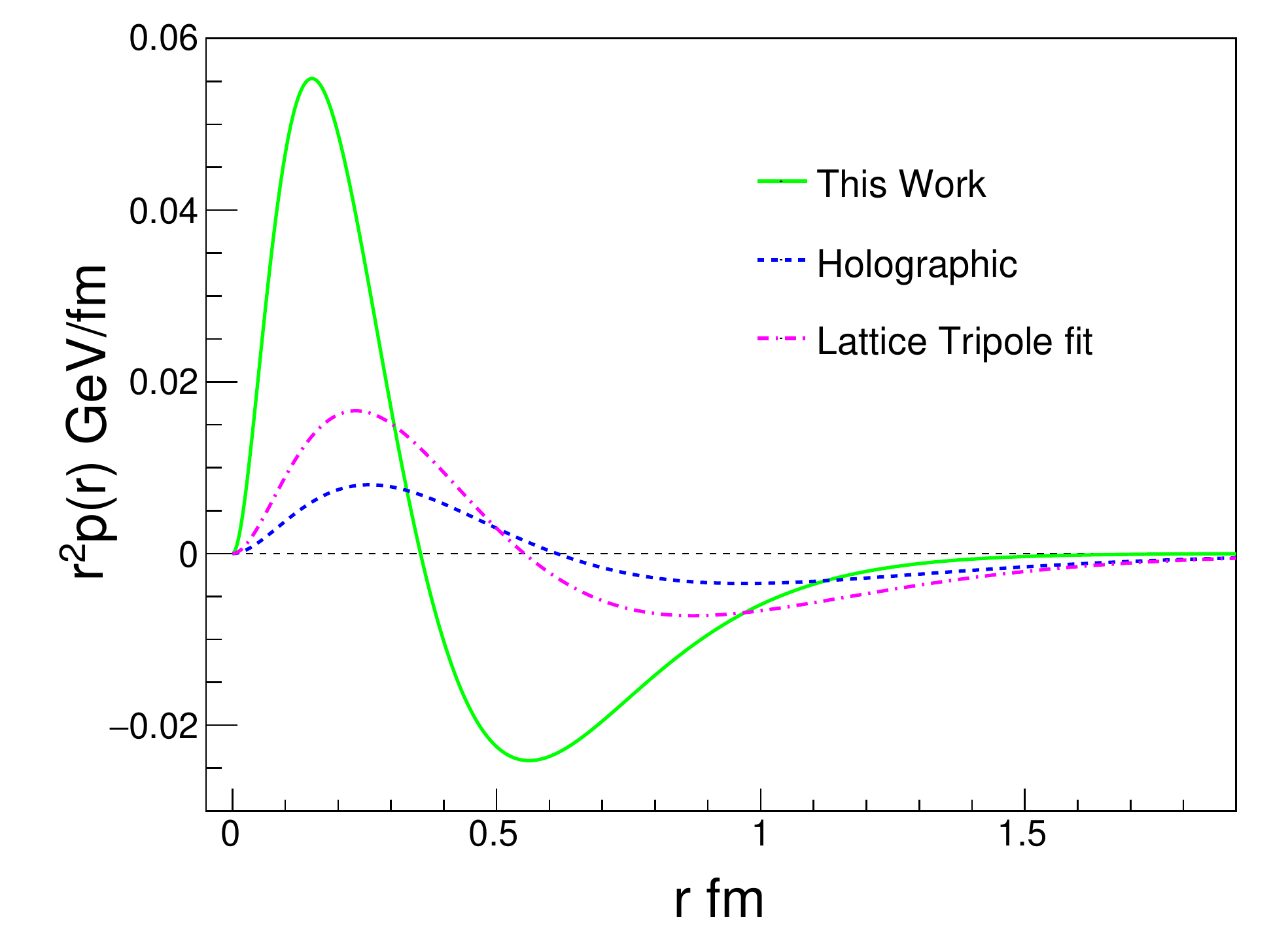}}
	\subfigure{
		\includegraphics[width=0.46\textwidth]{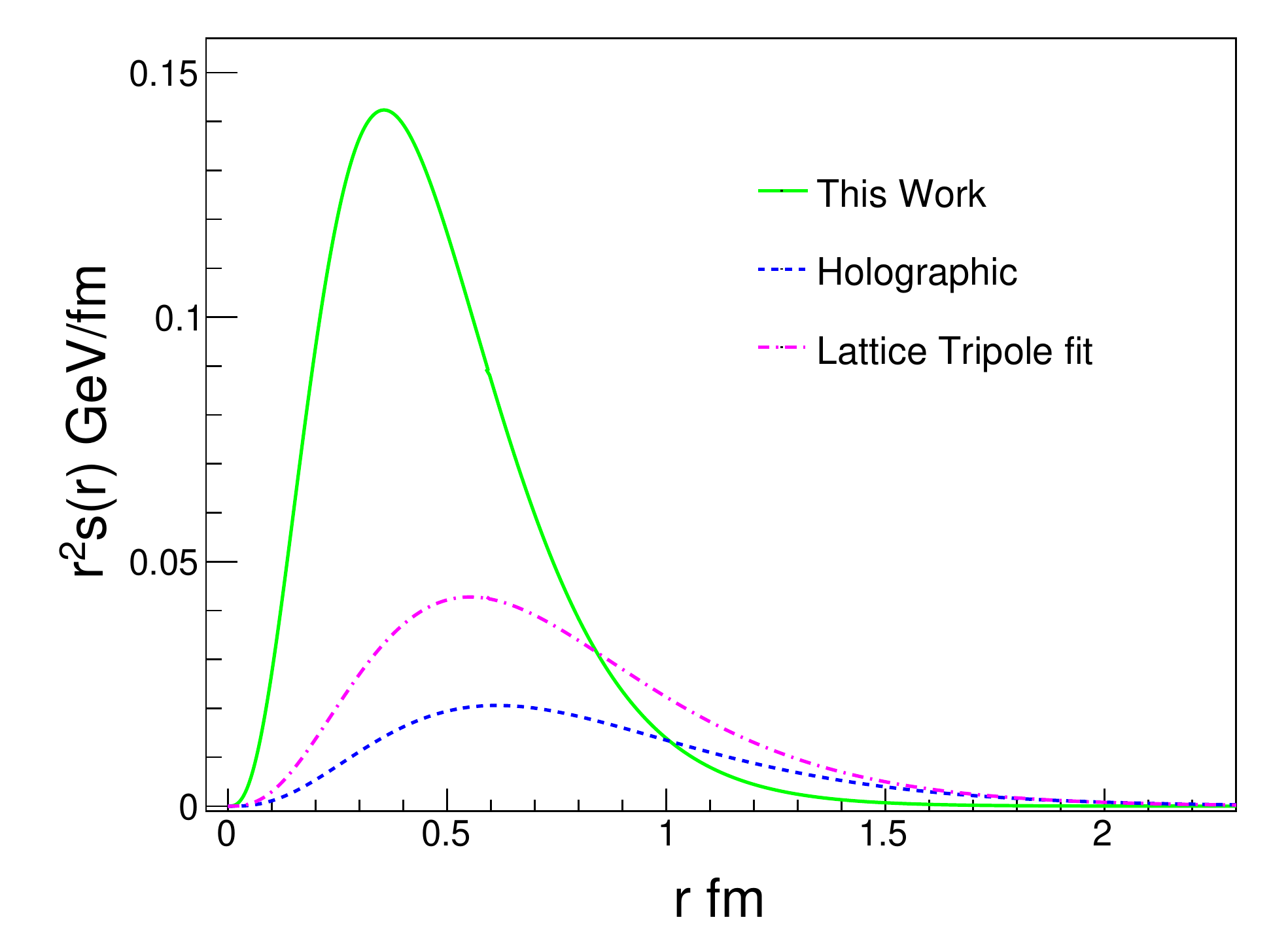}}
	\caption{(Left) The pressure distribution $r^2p(r)$ in GeV/fm inside proton. The green solid curve represents our result with CE method and the other two curves are from Refs. \cite{Mamo:2022eui} and \cite{Pefkou:2021fni} respectively.  (Right) The shear force distribution $r^2s(r)$ in GeV/fm inside proton. The green solid curve represents our result with CE method and the other two curves are from Refs. \cite{Mamo:2022eui} and \cite{Pefkou:2021fni} respectively.}
	\label{fig:pressure}
\end{figure*}

The radial pressure and shear distribution within protons exhibit reliance on the D-form factor, showcasing its discretization in FIG. \ref{fig:pressure}. Upon closer inspection of the left panel, each of the three curves contains at least one node whose position is determined by the Laue stability criterion. Notably, the outcomes obtained via the CE approach show higher results in both pressure and shear distribution compared to the other two methods. This is fundamentally due to the uniform handling of the A and D-form factors, an approach that differs from the individual treatment of form factors in the other two methods. Furthermore, the distributions achieved through CE are markedly compact relative to those obtained via alternative means, providing insights into the existence of dense regions inside the proton as mentioned in Ref. \cite{Duran:2022xag}.

In this letter, we provide a brief elaboration on the uncertainty analysis of our work. We note that the source of uncertainty in the mechanical quantities obtained, such as the proton radius and pressure, arises entirely from the parameterization of the form factor itself, as we do not consider any experimental data constraints on the parameters. Our calculations employ the approximation 
$D(K^2)=-4A(K^2)$, which is also documented in Ref. \cite{Mamo:2019mka}. Notably, different approaches have been adopted for determining $A(K^2)$ and $D(K^2)$ separately, as discussed in Ref. \cite{Mamo:2022eui}. However, we have deferred further exploration of its CE implementation to future work, given its complexity. The extraction uncertainty of the parameter $\lambda_N$ is primarily attributed to the CE method, with a small systematic uncertainty arising from numerical calculations. Moreover, we observe differences between our results for the radial pressure and shear distribution of the proton and those reported by other methods, mainly due to variations in the determination of the D-form factor, as detailed in the text.

\section{Conclusion and outlook}
\label{sec:summary}

In this letter, we introduce a novel perspective for characterizing the structure of the proton using the CE with the holographic QCD model. Specifically, we calculate the energy density of the system corresponding to the 00th component of the proton EMT and identify the corresponding CE critical point. We emphasize that 
 the energy density and the GFFs that make up it are modeled by holographic QCD. This critical point then determines the parameter $\lambda_N$ for the proton gravitational form factor. At present, the normalization factor for the gravitational form factors is still determined using lattice calculations. Using the obtained parameter, we determine the proton mass radius and scalar radius as defined by Eqs. (\ref{eq:Gs}-\ref{eq:radii}). Our calculated results yield 
 $\sqrt{\langle r_M^2\rangle}=0.720$ fm, $\sqrt{\langle r_S^2\rangle}=1.024$ fm for proton mass and scalar radius respectively, which are generally consistent with radii obtained from other methods. Additionally, we utilize the CE method to obtain the radial pressure distribution and shear force distribution inside the proton. Notably, these results are based on certain assumptions, and future work will aim to investigate the general case of determining the gravitational form factor using CE, without necessarily relying on the approximation $D(K^2)=-4A(K^2)$. 
 
 The use of CE as a method for studying nucleon structure within the framework of informational theory continues to be an area of active investigation, with ongoing efforts to systematically evaluate its validity and limitations. Future high-precision experiments involving $\Upsilon$ photoproduction near threshold at the electron-ion collider \cite{Accardi:2012qut,Chen:2018wyz,Chen:2020ijn,Anderle:2021wcy} will serve to constrain and test the parameters of current models.

\begin{acknowledgments}
We are grateful to Cédric Lorcé and  
Kazem Azizi for discussions and comments on the manuscript. This work is supported by the Strategic Priority Research Program of Chinese Academy of Sciences (Grant NO. XDB34030301).
\end{acknowledgments}

\bibliographystyle{apsrev4-1}
\bibliography{refs}
\newpage

\end{document}